\documentclass[twocolumn,npa,showpacs]{revtex4}
\usepackage{amsmath,bm}%
\usepackage{graphicx}%
\usepackage{ulem}

\begin{document}

\draft

\title{ Multistage Random Growing Small-World
Networks with Power-law degree Distribution
}

\author{\footnotesize LIU Jian-Guo, Dang Yan-Zhong, and Wang Zhong-Tuo\\
}
\address{Institute of System Engineering, Dalian University of Technology,
Dalian 116023, P. R. China}


\begin{abstract} \textnormal{\small {In this paper, a simply rule that generates scale-free networks
with very large clustering coefficient and very small average
distance is presented. These networks are called {\bf Multistage
Random Growing Networks}(MRGN) as the adding process of a new node
to the network is composed of two stages. The analytic results of
power-law exponent $\gamma=3$ and clustering coefficient $C=0.81$
are obtained, which agree with the simulation results
approximately. In addition, the average distance of the networks
increases logarithmical with the number of the network vertices is
proved analytically. Since many real-life networks are both
scale-free and small-world networks, MRGN may perform well in
mimicking reality.}}
\end{abstract}
\keywords{Complex networks, Scale-free networks, Small-world
networks, Disordered systems.}

\pacs{89.75.Da, 89.75.Fb, 89.75.Hc}

\maketitle

The past few years have witnessed a great devotion by physicists
to understand and characterize the underlying mechanisms of
complex networks including the Internet, the World Wide Web, the
scientific collaboration networks and so
on\cite{WS98,BA99,AB02,DM02,New,XFWang01}. The results of many
experiments and statistical analysis indicate that the networks in
various fields have some common characteristics. They have a small
average distance like random graphs, a large clustering
coefficient and power-law degree distribution \cite{WS98,BA99},
which is called the small-world and scale-free characteristics.
Recent works on the mathematics of networks have been driven
largely by the empirical properties of real-life
networks\cite{E1,E2,E3,E4,E5,E6,E7,E8,E9} and the studies on
network
dynamics\cite{D1,D2,D3,D4,D5,D6,D7,D8,D9,D10,D11,D12,D13,D14,D15},
optimization\cite{O1,O2,O3,O4,O5,O6} and evolutionary
\cite{M1,M2,M3,M4,M5,M6,M7,M8,M9,M10,M11,M12,M13}. The first
successful attempt to generate networks with high clustering
coefficients and small average distance is that of Watts and
Strogatz (WS model) \cite{WS98}. Another significant model is
proposed by Barab\'{a}si and Albert called scale-free network (BA
network) \cite{BA99}. The BA model suggests that growth and
preferential attachment are two main self-organization mechanisms
of the scale-free networks structure. These point to the fact that
many real-world networks continuously grow by the way that new
nodes and edges are added to the network, and new nodes would like
to attach to the existing nodes with large number of neighbors.

Dorogovtsev {\it et. al} proposed an simple model of scale-free
growing networks for any size of the network \cite{M6}. The idea
of the model is that a new node is added to the network at each
time step, which connects to both ends of a randomly chosen link
undirected. The model can be described by the process that the
newly added node connect to node $i$ preferentially, then select a
neighbor node of the node $i$ randomly. Holme {\it et. al}
proposed the famous model to generate growing scale-free networks
with tunable clustering \cite{M7}. The model introduced a
additional step to get the trial information and demonstrated that
the average number of triad formation trials controls the
clustering coefficient of the network. It should be noticed that
the newly added node connected the first node $i$ preferentially.
Actually, it would like to connect the neighbor nodes of node $i$
preferentially. Inspired by these questions, we give the
multistage random growing networks model. At each time step, the
new node is added to the network preferentially, then it would
find one of the node's neighbors to connect preferentially.

A scale-free small-world network using a very simple rule is
presented. The network starts with a triangle containing three
nodes marked as I, II and III. At each time step, a new node is
added to the network with two edges. The first edge would choosing
node to connected depends on the degree $k_i$ of node $i$, such
that $k_i/\sum k_i$, and then attach another edge to a node which
is connected with the first selected node preferentially.
According to this process, the general iterative algorithm of MRGN
is introduced. $A(t)$ denotes MRGN after $t$ iterations. Since the
network size increases by one at each time step, $t$ is used to
represent the node added in the $t$th step. At step $t$, we can
easily see that the network consists of $N=t+3$ vertices. The
total degree equals $4t+3$. When $t$ is large, the average degree
at step $t$ is equal approximate to a constant value $4$, which
shows that MRGN is sparse like many real-life network
\cite{AB02,DM02,New}.
The topology characteristics of the model are analyzed both
analytically and by numerical calculations. The analytical
expressions agree with the numerical simulations approximately.



The distribution is one of the most important statistical
characteristics of networks. Since many real-world networks are
scale-free networks, whether the network is of the power-law
degree distribution is a criterion to judge the validity of the
model. By using the mean-field theory, the evolution of the degree
distribution of individual nodes can be described as following
\begin{equation}\label{F2.1}
\frac{\partial k_i}{\partial t}=P(i)+\sum_{j\in
\Gamma_i}P(i|j)P(j),
\end{equation}
where $P(i)$ denotes the possibility that the node $i$ with degree
$k_i$ is selected in the first step, $P(i|j)$ denotes the
conditional possibility that node $i$ is the neighbor of node $j$
with degree $k_j$ which have been selected at the first step and
$\Gamma_i$ denotes the neighbor node set of node $i$. Because the
new node is added to the network preferentially, one has
\begin{equation}\label{F2.2}
P(i)=\frac{k_i}{\sum_{j=1}^{N-1} k_j}.
\end{equation}
The conditional possibility $P(i|j)$ can be calculated by
\begin{equation}\label{F2.3a}
P(i|j)=\frac{k_i}{\sum_{l\in \Gamma_j}k_l}.
\end{equation}
Since every newly added node has two edges, $P(i|j)$ can be
approximately by
$P(i|j)=\frac{k_i}{k_l\langle k\rangle}.$
Then, one can get that
\begin{equation}\label{F2.5}
\frac{\partial k_i}{\partial t}=\frac{k_i}{\sum_j k_j}+\sum_{j\in
\Gamma_i}\frac{k_i}{\langle k\rangle k_l}\frac{k_l}{\sum_j
k_j}=\frac{k_i}{2\sum_j k_j}.
\end{equation}
The sum in the denominator goes over all nodes in the network
except the newly introduced one, thus its value is $\sum_j
k_j=2t+3$. The solution of Equ. (\ref{F2.5}), with the initial
condition that every node $i$ at its introduction has
$k_i(t_i)=2$, is
\begin{equation}\label{F2.6}
k_i(t)=(\frac{t}{t_i})^{\beta},
\end{equation}
where $\beta=0.5$. One can get that the degree distribution of
MRGN is as following
\begin{equation}\label{F2.7}
P(k)\sim k^{-\gamma},
\end{equation}
where $\gamma=\frac{1}{\beta}+1=3$. The numerical simulation
results are demonstrated in Fig. 1.

\begin{figure}[ht]
  \begin{center}
       \center \includegraphics[width=8cm]{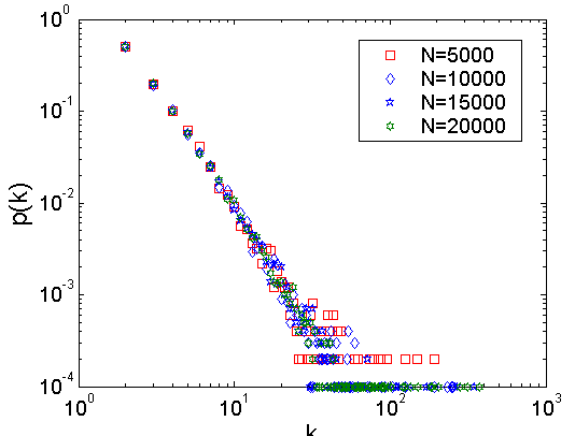}
       \caption{Degree distribution of MRGN, with $N=20000$ (hexagons),
       $N=15000$ (pentagons), $N=10000$ (diamonds) and $N=5000$ (squares).
       In this figure, $p(k)$ denotes the probability of the number of nodes with
       degree $k$ in the network. The power-law degree distribution exponent $\gamma$ of the
       four probability density function are $\gamma_{25000}=2.88\pm 0.02$,
       $\gamma_{20000}=2.88\pm 0.05$, $\gamma_{15000}=2.86\pm 0.06$ and $\gamma_{10000}=2.85\pm 0.02$}
 \end{center}
\end{figure}


As we have mentioned above, the degree distribution is one of the
most important statistical characteristics of networks. The
average distance is also one of the most important parameters to
measure the efficiency of communication network. The average
distance $L$ of the network is defined as the mean distance over
all pairs of nodes.  The average distance plays a significant role
in measuring the transmission delay. Marked each node of the
network according to the time when the node is added to the
network. Firstly, we give the following lemma \cite{M1}.

{\bf Lemma 1} {\it For any two nodes $i$ and $j$, each shortest
path from $i$ to $j$ does not pass through any nodes $k$
satisfying that $k>{\rm max}\{i,j\}$.
}

{\it Proof.} Denote the shortest path from node $i$ to $j$ of length
$n+1$ by $i\rightarrow x^1 \rightarrow x^2 \cdots \rightarrow x^n
\rightarrow j$($SP_{ij}$),  where $n>0$. Suppose that $x^k={\rm
max}\{x^1, x^2, \cdots, x^n\}$, if $k\leq {\rm max}\{i, j\}$, then
the conclusion is true.

Then we prove the case that $k>{\rm max}\{i,j\}$ would not come
forth. Suppose the edge $E_{y_1y_2}$ is selected when node $x_k$ is
added. If $k>{\rm max}\{i,j\}$, neither node $i$ nor node $j$ is
belong to the $E_{y_1y_2}$. Hence the path from $i$ to $j$ passing
through $x^k$ must enter and leave $E_{y_1y_2}$. Assume that the
path enter $E_{y_1y_2}$ by node $y_1$ and leave from node $y_2$,
then there exists a path of $SP_{ij}$ from $y_1$ to $y_2$ passing
through $x^k$, which is longer than the direct path $y_1\rightarrow
y_2$. The youngest node must be either $i$ or $j$ when $SP_{ij}$ is
the shortest path. 

Denote $d(i,j)$ as the distance between node $i$ and node $j$. Let
$\sigma(N)$ represent the total distance $\sigma(N)=\sum_{1\leq
i<j\leq N}d(i,j)$. The average distance of MRGN with order $N$,
denoted by $L(N)$, is defined as following
\begin{equation}\label{F3.1}
L(N)=\frac{2\sigma(N)}{N(N-1)}.
\end{equation}
According to Lemma 3.1, the node newly added in the network will not
affect the distance between old nodes. Hence we have
\begin{equation}\label{F3.2}
\sigma(N+1)=\sigma(N)+\sum^N_{i=1}d(i,N+1).
\end{equation}
Assume that the $(N+1)$th node is add to the edge $E_{y_1y_2}$, then
Equ.(\ref{F3.2}) can be written as
\begin{equation}\label{F3.3}
\sigma(N+1)=\sigma(N)+N+\sum^N_{i=1}D(i,y).
\end{equation}
where $D(i,y)={\rm min}\{d(i,y_1),d(i,y_2)\}$. Let a single node $y$
represent the $E_{y_1y_2}$ continuously, then we have the following
equation
\begin{equation}\label{F3.4}
\sigma(N+1)=\sigma(N)+N+\sum_{i=\Lambda}d(i,y),
\end{equation}
where the node set $\Lambda=\{1,2, \cdots, N\}-\{y_1, y_2\}$ have
$(N-2)$ members. The sum $\sum_{i=\Lambda}d(i,y)$ can be considered
as the distance from each node of the network to node $y$ in MRGN
with order $N-1$. Approximately, the sum $\sum_{i=\Lambda}d(i,y)$ is
equal to $L(N-1)$. Hence we have
\begin{equation}\label{F3.5}
\sum_{i=\Lambda}d(i,y)\approx (N-2)L(N-1)
\end{equation}
Because the average distance $L(N)$ increases monotonously with $N$,
this yields
\begin{equation}\label{F3.6}
(N-2)L(N-1)=(N-2)\frac{2\sigma(N-1)}{(N-1)(N-2)}<\frac{2\sigma(N)}{N-1}.
\end{equation}
Then we can obtain the inequality
\begin{equation}\label{F3.7}
\sigma(N+1)<\sigma(N)+N+\frac{2\sigma(N)}{N-1}.
\end{equation}
Enlarge $\sigma(N)$, then the upper bound of the increasing tendency
of $\sigma(N)$ will be obtained by the following equation
\begin{equation}\label{F3.8}
\frac{d\sigma(N)}{dN}=N+\frac{2\sigma(N)}{N-1}.
\end{equation}
This leads to the following solution
\begin{equation}\label{F3.9}
\sigma(N)={\rm log}(N-1)(N-1)^2+C_1(N-1)^2-(N-1).
\end{equation}

\begin{figure}[ht]
  \begin{center}
       \center \includegraphics[width=8cm]{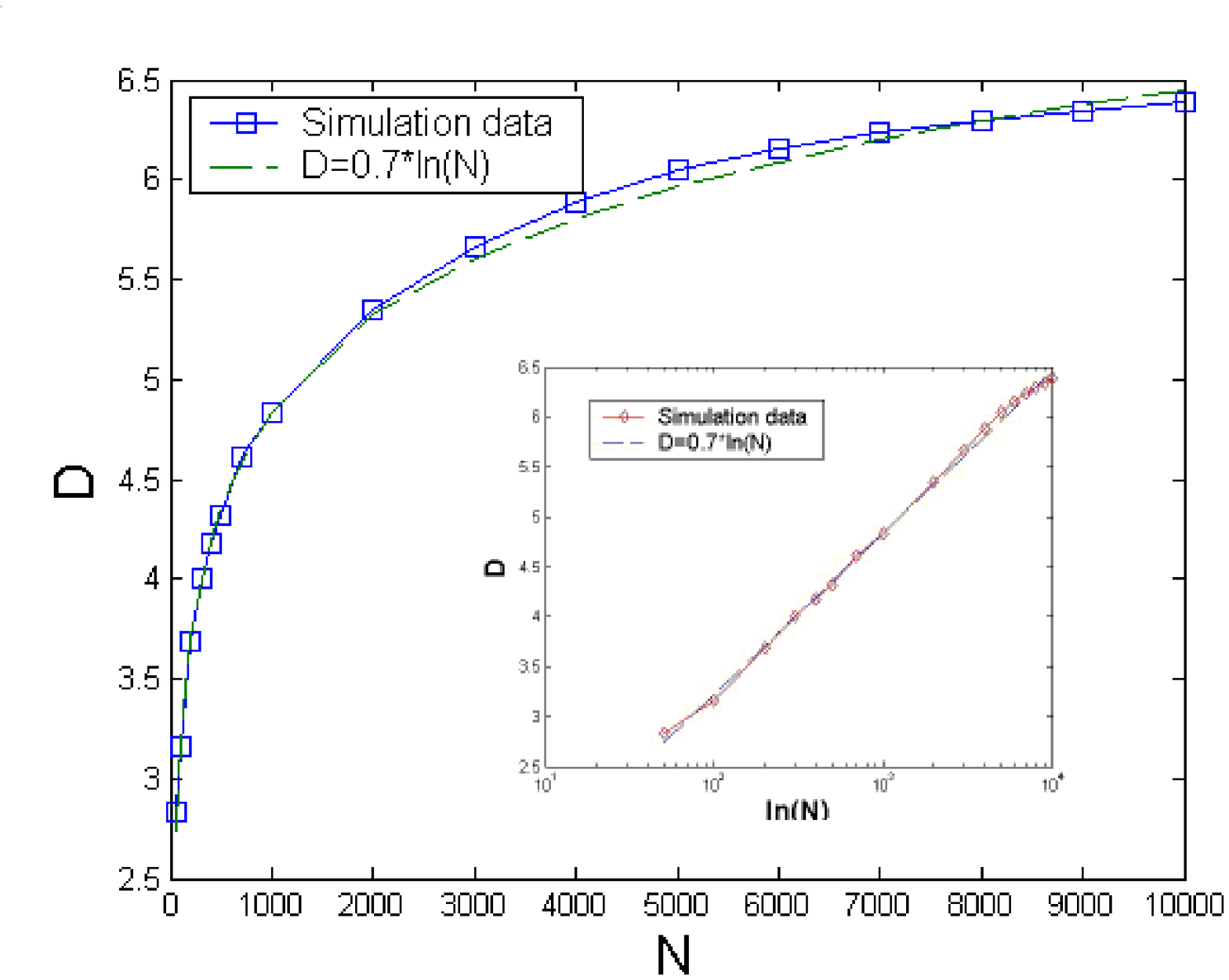}
       \caption{The dependence between the average distance $L$ and
the order $N$ of MRGN. One can see that $L$ increases very slowly as
$N$ increases. The inset exhibits the curve where $L$ is considered
as a function of ${\rm ln}N$, which is fitted by a straight line.
The curve is above the fitting line when $N$ is small($2000\leq
N\leq 7000$) and under the fitting line when $N$ is large($N\geq
8000$), which indicates that the increasing tendency of $L$ can be
approximated as ${\rm ln}N$ and in fact a little slower than ${\rm
ln}N$. All the data are obtained by 10 independent simulations.}
 \end{center}
\end{figure}

\begin{figure}[ht]
  \begin{center}
       \includegraphics[width=0.4\textwidth]{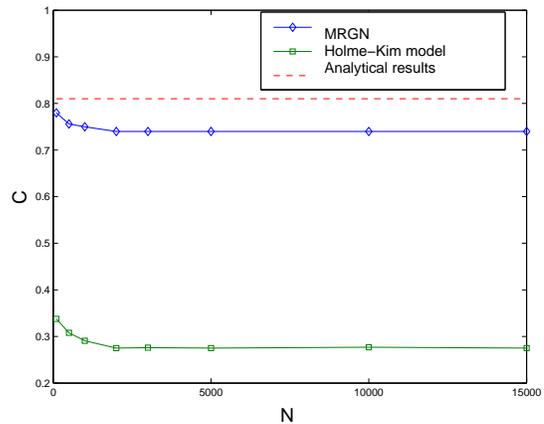}
       \caption{The clustering coefficient of MRGN(red diamonds) and Holme-Kim networks(green squares). In
       this figure, one can find that the clustering coefficient of
       MRGN is almost a constant a little smaller than 0.75. The red line represents
       the analytic result 0.81. It is clear that, the clustering
       coefficient of Holme-Kim networks is much smaller than that of MRGN.
       }
 \end{center}
\end{figure}

By means of the theoretic approximate calculation, we prove that the
increasing tendency of $L(N)$ is a little slower than ln$N$. In Fig
3, we report the simulation results on average distance of MRGN,
which agree with the analytic result.

The small-world effect consists of two properties: large
clustering coefficient and small average distance. The clustering
coefficient, denoted by $C$, is defined as $
C=\sum_{i=1}^N\frac{C_i}{N}$, where $C_i$ is the clustering
coefficient for any arbitrary node $i$. $C_i$ is
\begin{equation}\label{F4.1}
C_i=\frac{2E(i)}{k_i(k_i-1)},
\end{equation}
where $E(i)$ is the number of edges in the neighbor set of the node
$i$, and $k_i$ is the degree of node $i$. When the node $i$ is added
to the network, it is of degree 2 and $E(i)=1$. If a new node is
added to be a neighbor of $i$ at some time step, $E(i)$ will
increase by one since the newly added node will link to one of the
neighbors of node $x$. Therefore, in terms of $k_i$ the expression
of $E(i)$ can be written as following
\begin{equation}\label{F4.2}
E(i)=1+(k_i-2)=k_i-1.
\end{equation}
Hence, we have that
\begin{equation}\label{F4.3}
C_i=\frac{2(k_i-1)}{k_i(k_i-1)}=\frac{2}{k_i}.
\end{equation}
This expression indicates that the local clustering scales as
$C_i\sim k^{-1}$. It is interesting that a similar scaling has
been observed in pseudofractal web \cite{M8} and several real-life
networks \cite{M9}. Consequently, we have
\begin{equation}\label{F4.4}
C=\frac{2}{N}\sum^N_{i=1}\frac{1}{k_i}.
\end{equation}
Since the degree distribution is $p(k)=c_1 k^{-3}$, where
$k=2,3,\cdots, k_{\rm max}$. The average clustering coefficient $C$
can be rewritten as
\begin{equation}
C=\sum_{k=2}^{k_{\rm max
}}\frac{2}{N}\frac{Np(k)}{k}=2c_1\sum_{k=2}^{k_{\rm max}} k^{-4}.
\end{equation}
For sufficient large $N$, $k_{max}\gg 2$. The parameter $c_1$
satisfies the normalization equation
\begin{equation}\label{F4.8}
\sum_{k=2}^{k_{\rm max}}p(k)dk=1.
\end{equation}

It can be obtain that $c_1=4.9491$ and $C=2\times
4.9491\times\sum_{k=2}^{k_{\rm max}}k^{-4}=0.8149$. From Fig. 4,
we can get that the analytical average clustering coefficient
deviance the real value a little. Because the analytic one is
obtained when the time step $t\rightarrow \infty$ and the
simulation result is obtained when the time step $t$ is finite.
The other reason is that simulation result $\gamma$ of the degree
distribution deviant $3$ a little, which is caused the finite
network size. However, the most important reason lies in the
hypothesis (\ref{F2.5}) that there are no correlations between all
nodes. The demonstration exhibits that most real-life networks
have large clustering coefficient no matter how many nodes they
have. That is agree with the case of MRGN but conflict with the
case of BA networks, thus MRGN may be more appropriate to mimic
the reality.

In summary, we have introduced a simple iterative algorithm for
constructing MRGN. The networks have very large clustering
coefficients and very small average distance, which satisfy many
real networks characteristics, such as the technological and
social networks. After the newly added node connect to the first
node $i$, it connect to the neighbor node of node $i$
preferentially. They are not only the scale-free networks, but
also small-world networks. The results imply the following
conclusion: if there are no correlation between all node and the
new node adds to the network in two step, whether the second step
is random or preferential, the degree distribution would be
power-law and the exponent is 3. We have computed the analytical
expressions for the degree distribution and clustering
coefficient. Since most real-life networks are both scale-free and
small-world networks, MRGN may perform better in mimicking
reality. Further work should focus on the information flow and the
epidemic spread on MRGN.

This work has been supported by the Chinese Natural Science
Foundation of China under Grant Nos. 70431001 and 70271046.

\end{document}